\documentclass{article}
\usepackage{amssymb}
\usepackage{amsmath}
\begin{document}
\baselineskip16pt
\title{Brownian motion in a magnetic field }
\author{Rados{\l}aw Czopnik \\
Institute of Theoretical Physics, University of Wroc{\l}aw, \\
PL-50 205 Wroc{\l}aw, Poland\\ and \\
 Piotr Garbaczewski\\ Institute of
Physics, Pedagogical University,\\ PL-65 069 Zielona G\'{o}ra,
Poland}

\maketitle

\begin{abstract}
We derive explicit forms  of Markovian transition
probability densities for  the velocity space,  phase-space
and the Smoluchowski configuration-space Brownian motion of a charged
particle in a constant magnetic field. By invoking a
hydrodynamical formalism for those stochastic processes, we
quantify a continual (net on the local average) heat transfer
from the thermostat to diffusing particles.
 \end{abstract}

\section{\protect\bigskip Introduction}

We address an old-fashioned problem of  the Brownian motion of a
charged particle in a constant magnetic field. That issue has
originated from studies of the diffusion of plasma across a
magnetic field \cite{Tay}, \cite{Kur} and nowadays, together with
a free Brownian motion example,  stands for a textbook
illustration of how  transport and auto-correlation functions
should be computed in generic situations governed by the Langevin
equation (to a suitable degree of approximation of a kinetic
theory, when collisions are stochastically modeled in terms  of a
random force), cf. \cite{Bal} but also \cite{Sch}, \cite{vKa}.

From a purely pragmatic point of view   this white-noise strategy
is quite  satisfactory. After
(formally) evaluating velocity auto-correlation functions,
 formulas for running and asymptotic diffusion coefficients
easily follow. To that end  an explicit form of the
probability density or transition probability density of the
involved stochastic diffusion processes (in velocity space,
phase-space or configuration space)  is not necessary, cf. \cite{Tay},
\cite{Bal}.

To our knowledge, except for the paper \cite{Kur} (mentioned in
\cite{vKa} as a footnote reference for the purpose of evaluation
of the mean square velocity and its mean square displacement at
equilibrium), for a Brownian particle in a constant magnetic field
no attempt was made in the literature to  give a complete
characterization of the stochastic  process itself, nor pass to
the associated macrosopic (hydrodynamical formalism) balance
equations.  (Cf. \cite{Gar}, \cite{Gar1}, \cite{Gar2},
\cite{Step}, \cite{Tit},
 \cite{Lag} for a  number of reasons why to do that).

Surprisingly enough, in Ref. \cite{Kur},
 the Brownian motion in a magnetic field
 is described in terms of \it
operator-valued \rm (matrix-valued functions) probability
distributions that additionally involve fractional powers of
matrices. In consequence, there is no clean path  towards a
(necessary) relationship with  the associated Kramers-Smoluchowski
equations (cf. Chap. 6.1 in Ref. \cite{Sch}), nor ways to stay in
conformity  with  the standard wisdom about probabilistic
procedures valid in case of the free Brownian motion
(Ornstein-Uhlenbeck process), cf. \cite{Cha}, \cite{Nel},
\cite{Step}.

Therefore, we decided to address an issue of the Brownian motion
in a magnetic field anew, to unravel its features  of
a fully-fledged stochastic diffusion process.
In particular,
we  derive transition probability densities governing both the velocity,
phase-space and the configuration space processes.
Hydrodynamical  balance equations and their behaviour in the
Smoluchowski regime  are discussed as well.

\section{Velocity-space diffusion process}

The standard analysis of the Brownian motion of a free particle employs
the Langevin equation
$\frac{d\overrightarrow{u}}{dt}=-\beta \overrightarrow{u}+
\overrightarrow{A}%
\left( t\right)$ where $\overrightarrow{u}$ denotes the velocity
of the particle and  the influence of the surrounding medium on
the motion (random acceleration) of the particle is modeled by
means of two independent contributions. A systematic part $-\beta
\overrightarrow{u}$ represents  a dynamical friction. The
remaining fluctuating part $\overrightarrow{A}\left( t\right) $ is
supposed to display a statistics of  the familiar white noise:
(i)$\overrightarrow{A}\left( t\right) $ is independent of $%
\overrightarrow{u}$, (ii) $\left\langle A_{i}\left( s\right)
\right\rangle = 0$ and $\left\langle A_{i}\left( s\right)
A_{j}\left( s^{\shortmid }\right) \right\rangle =2q\delta
_{ij}\delta \left( s-s^{\shortmid }\right) $ for $i,j=1,2,3$,
where $q=\frac{k_{B}T}{m}\beta $ is a physical parameter. The
well-known Ornstein-Uhlenbeck stochastic process comes out
 on that conceptual basis, \cite{Cha},\cite{Nel}, \cite{Step}.

The linear friction model  can be  adopted  to the case of
diffusion of charged particles in the presence of a constant
magnetic field which acts upon particles via the Lorentz force.
The Langevin equation for that  motion reads:

\begin{equation}
\frac{d\overrightarrow{u}}{dt}=-\beta \overrightarrow{u}+\frac{q_{e}}{mc}%
\overrightarrow{u}\times \overrightarrow{B}+\overrightarrow{A}\left( t\right)
\label{Langevin}
\end{equation}

where $q_{e}$ denotes an  electric charge of the particle of mass $m$.

Let us assume for simplicity that the constant magnetic field
$\overrightarrow{B}$ is
directed along the z-axis of a Cartesian reference frame:
$ \overrightarrow{B}=\left( 0,0,B\right) $ and $B=const$. In this case
Eq. (\ref{Langevin}) takes the form

\begin{equation}
\frac{d\overrightarrow{u}}{dt}=-\Lambda \overrightarrow{u}+
\overrightarrow{A}%
\left( t\right)  \label{LanII}
\end{equation}

where
\begin{equation}
\Lambda =\left(
\begin{array}{ccc}
\beta  & -\omega _{c} & 0 \\
\omega _{c} & \beta  & 0 \\
0 & 0 & \beta
\end{array}
\right)
\end{equation}

 and $\omega _{c}=\frac{q_{e}B}{mc}$
 denotes the Larmor frequency.
Assuming the Langevin equation to be (at least formally) solvable,
we can infer  a probability density  $P\left( \overrightarrow{u},t|%
\overrightarrow{u}_{0}\right) $, $t>0$ conditioned by the  the initial
velocity data choice $\overrightarrow{%
u}=\overrightarrow{u}_{0}$ at $t=0$. Physical circumstances of the
problem enforce a demand: (i) $P\left(
\overrightarrow{u},t|\overrightarrow{u}_{0}\right)
\rightarrow \delta ^{3}\left( \overrightarrow{u}-\overrightarrow{u}%
_{0}\right) $ as $t\rightarrow 0$ and (ii)  $P\left(
\overrightarrow{u},t|\overrightarrow{u}_{0}\right) \rightarrow
\left( \frac{m}{2\pi k_{B}T}\right) ^{\frac{3}{2}}\exp \left(
-\frac{m |\overrightarrow{u}_{0}| ^{2}}{2k_{B}T} \right) $ as
$t\rightarrow \infty $.

A  formal solution of Eq. (\ref{LanII}) reads:

\begin{equation}
\overrightarrow{u}\left( t\right) -e^{-\Lambda t}\overrightarrow{u}%
_{0}=\int_{0}^{t}e^{-\Lambda \left( t-s\right) }
\overrightarrow{A}\left( s\right) ds \enspace .  \label{sol1}
\end{equation}

By taking into account that
\begin{equation}
e^{-\Lambda t}=e^{-\beta t}\left(
\begin{array}{ccc}
\cos \omega _{c}t & \sin \omega _{c}t & 0 \\
-\sin \omega _{c}t & \cos \omega _{c}t & 0 \\
0 & 0 & 1
\end{array}
\right) =e^{-\beta t}U\left( t\right)
\end{equation}

we can rewrite (\ref{sol1}) as  follows

\begin{equation}
\overrightarrow{u}\left( t\right) -e^{-\beta t}U\left( t\right)
\overrightarrow{u}_{0}=\int_{0}^{t}e^{-\beta \left( t-s\right)
}U\left( t-s\right) \overrightarrow{A}\left( s\right) ds  \enspace
.
\end{equation}

Statistical properties of $\overrightarrow{u}\left( t\right)
-e^{-\Lambda t}\overrightarrow{u}_{0}$ are identical with those of
 $% \int_{0}^{t}e^{-\Lambda \left( t-s\right) }
\overrightarrow{A}\left( s\right)
ds$.
 In consequence, the problem of deducing a probability density
 $P\left(
\overrightarrow{u},t|\overrightarrow{u}_{0}\right) $ is
equivalent to deriving  the probability distribution
of the random vector

\begin{equation}
\overrightarrow{S}=\int_{0}^{t}\psi
\left( s\right) \overrightarrow{A}\left(
s\right) ds  \label{Sdef}
\end{equation}

where  $\psi \left( s\right) =e^{-\Lambda \left( t-s\right)
}=e^{-\beta \left( t-s\right) }U\left( t-s\right)$.

The white noise term $ \overrightarrow{A}\left( s\right) $ in view
of the integration with respect to time is amenable to a more
rigorous analysis that invokes the Wiener process increments and
their statistics, \cite{Doob}. Let us divide the time integration
interval into a  large number of small subintervals  $\Delta t$.
We adjust them suitably to assure  that effectively $\psi \left(
s\right) $ is constant on each subinterval $\left( j\Delta
t,\left( j+1\right) \Delta t\right) $ and equal $\psi \left(
j\Delta t\right) $. As a result we obtain the expression

\begin{equation}
\overrightarrow{S}=\sum_{j=0}^{N-1}\psi \left( j\Delta t\right)
\int_{j\Delta t}^{\left( j+1\right) \Delta
t}\overrightarrow{A}\left( s\right) ds \enspace .  \label{S}
\end{equation}

Here $ \overrightarrow{B}\left( \Delta t\right) =\int_{j\Delta
t}^{\left( j+1\right) \Delta t}\overrightarrow{A}\left( s\right)
ds$ stands for the above-mentioned Wiener process  increment.
Physically, $\overrightarrow{B}\left( \Delta t\right) $
 represents the \it net \rm acceleration which a Brownian particle
may suffer (in fact accumulates)  during an interval of
time $\Delta t$.

Equation (\ref{S}) becomes

\begin{equation}
\overrightarrow{S}=\sum_{j=0}^{N-1}\psi \left( j\Delta t\right)
\overrightarrow{B}\left( \Delta t\right) =
\sum_{j=0}^{N-1}\overrightarrow{s}%
_{j}
\end{equation}

where we introduce $\overrightarrow{s}_{j}=\psi \left( j\Delta t\right)
\overrightarrow{B}\left( \Delta t\right) =
\psi _{j}\overrightarrow{B}\left(
\Delta t\right) $.

The Wiener process argument \cite{Cha}, \cite{Nel}, \cite{Gar}
allows us to infer  the probability distribution of
$\overrightarrow{s}_{j}$. It is enough  to employ the fact that
the  distribution of $\overrightarrow{B}\left( \Delta t\right) $
is Gaussian with mean zero and variance $q=\frac{k_{B}T}{m}\beta
$. Then

\begin{equation}
w\left[ \overrightarrow{B}\left( \Delta t\right) \right] =
\left( \frac{1}{%
4\pi q\Delta t}\right) ^{\frac{3}{2}}\exp \left( -\frac{\left|
\overrightarrow{B}\left(
\Delta t\right) \right| ^{2}}{4q\Delta t}\right)
\label{w(B)}
\end{equation}

and in view of $\overrightarrow{s}_{j}=\psi _{j}
\overrightarrow{B}\left( \Delta t\right) $ by performing the
change of variables in (\ref{w(B)}) we get

\begin{equation}
\widetilde{w}\left[ \overrightarrow{s}_{j}\right] =
\det \left[ \psi _{j}^{-1}%
\right] w\left[ \psi _{j}^{-1}\overrightarrow{s}_{j}\right] =
\frac{1}{\det \psi _{j}}w\left[ \psi
_{j}^{-1}\overrightarrow{s}_{j}\right] \enspace .
\end{equation}

Since $
\det \psi \left( s\right) =e^{-3\beta \left( t-s\right) } $ and
$\psi ^{-1}\left( s\right) = U\left[ -\left( t-s\right) \right]
e^{\beta \left( t-s\right) }$  we obtain

\begin{equation}
\widetilde{w}\left[ \overrightarrow{s}_{j}\right] =\left( \frac{1}{4\pi
q\Delta t}\right) ^{\frac{3}{2}}\frac{1}{e^{-3\beta \left( t-j\Delta
t\right) }}\exp \left( -\frac{\left|
e^{\beta \left( t-j\Delta t\right) }U%
\left[ -\left( t-j\Delta t\right) \right]
\overrightarrow{s}_{j}\right| ^{2}%
}{4q\Delta t}\right)
\end{equation}

and finally

\begin{equation}
\widetilde{w}\left[ \overrightarrow{s}_{j}\right] =
\left( \frac{1}{4\pi
q\Delta t}\frac{1}{e^{-2\beta \left( t-j\Delta t\right) }}
\right) ^{\frac{3}{%
2}}\exp \left( -\frac{\left| \overrightarrow{s}_{j} \right|
^{2}}{4q\Delta te^{-2\beta \left( t-j\Delta t\right) }}\right)
\enspace .
\end{equation}

Clearly, $\overrightarrow{s}_{j}$ are
mutually independent random variables whose distribution is Gaussian
with mean zero and variance $\sigma
_{j}^{2}=2q\Delta te^{-2\beta \left( t-j\Delta t\right) }$.
Hence, the probability distribution of $\overrightarrow{S}=
\sum_{j=0}^{N-1}%
\overrightarrow{s}_{j}$  is again  Gaussian with mean zero. Its
variance equals  the sum of variances of $\overrightarrow{s}_{j}$
i.e.  $ \sigma ^{2}=\sum_{j}\sigma _{j}^{2}=2q\sum_{j}\Delta
te^{-2\beta \left( t-j\Delta t\right) }$.

Taking the limit $N\rightarrow \infty $ $(\Delta t\rightarrow 0)$ we
arrive at

\begin{equation}
\sigma ^{2}=2q\int_{0}^{t}dse^{-2\beta \left( t-s\right) }=
\frac{k_{B}T}{m}%
\left( 1-e^{-2\beta t}\right) \enspace .
\end{equation}

Because of  $\overrightarrow{S}=\overrightarrow{u}
\left( t\right) -e^{-\Lambda t}%
\overrightarrow{u}_{0}$ the transition probability density of the
Brownian particle velocity, conditioned by the initial data
$\overrightarrow{u}_0$  at $t_0=0$ reads

\begin{equation}
P\left( \overrightarrow{u},t|\overrightarrow{u}_{0}\right) =
\left( \frac{1}{%
2\pi \frac{k_{B}T}{m}\left( 1-e^{-2\beta t}\right) }
\right) ^{\frac{3}{2}%
}\exp \left( -\frac{\left| \overrightarrow{u}-
e^{-\Lambda t}\overrightarrow{u%
}_{0}\right| ^{2}}{2\frac{k_{B}T}{m}\left( 1-e^{-2\beta t}\right)
} \right) \enspace .
\end{equation}

The process is Markovian and time-homogeneous, hence the above
formula can be trivially extended to encompass the case of
arbitrary $t_{0}\neq 0$ : $ P\left(
\overrightarrow{u},t|\overrightarrow{u}_{0},t_{0}\right)$ arises
by substituting everywhere $t-t_0$ instead of $t$.

Physical arguments (cf. demand (ii) preceding Eq. (4)) refer to an
asymptotic probability distribution (invariant measure density)
$P(u)$  of the random variable
 $\overrightarrow{u}$  in the  Maxwell-Boltzmann form

\begin{equation}
P\left( \overrightarrow{u}\right) =\left( \frac{m}{2\pi k_{B}T}
\right) ^{\frac{3}{2}}\exp \left( -\frac{m\left|
\overrightarrow{u}\right| ^{2}}{2k_{B}T}\right) \enspace .
\end{equation}

This time-independent  probability density  together with the
time-homogeneous transition density (15) uniquely determine a
stationary Markovian stochastic process for which we can evaluate
various mean values.

Expectation values of velocity components vanish: $ \left\langle
u_{i}\left( t\right) \right\rangle = \int_{-\infty }^{\infty
}u_{i}P\left( \overrightarrow{u}\right) d\overrightarrow{u}=0 $
for $i=1,2,3$. The matrix of the second moments (velocity
auto-correlation functions) reads

\begin{equation}
 \left\langle
u_{i}\left( t\right) u_{j}\left( t_{0}\right) \right\rangle
=\int_{-\infty }^{\infty }u_{i}u_{j}^{0}P\left( \overrightarrow{u},t;%
\overrightarrow{u}_{0},t_{0}\right)
d\overrightarrow{u}d\overrightarrow{u}%
_{0}
\end{equation}

 where $i,j=1,2,3$ and in view of $ P\left(
\overrightarrow{u},t;\overrightarrow{u}_{0},t_{0}\right) = P\left(
\overrightarrow{u},t|\overrightarrow{u}_{0},t_{0}\right) P\left(
\overrightarrow{u}_{0}\right)$ we arrive at the compact expression

\begin{equation}
\frac{k_{B}T}{m}e^{-\Lambda \left| t-t_{0}\right| }=\frac{k_{B}T}{m}%
e^{-\beta \left| t-t_{0}\right| }\left(
\begin{array}{ccc}
\cos \omega _{c}\left| t-t_{0}\right| & \sin
\omega _{c}\left| t-t_{0}\right|
& 0 \\
-\sin \omega _{c}\left| t-t_{0}\right| & \cos \omega _{c}\left|
t-t_{0}\right| & 0 \\
0 & 0 & 1
\end{array}
\right) \enspace .
\end{equation}

In particular, the auto-correlation function (second moment) of
the $x$-component of  velocity equals
\begin{equation}
\left\langle u_{1}\left( t\right) u_{1}\left( t_{0}\right) \right\rangle =%
\frac{k_{B}T}{m}e^{-\beta \left| t-t_{0}\right| }\cos \omega _{c}\left|
t-t_{0}\right|   \label{autocor}
\end{equation}

in  agreement with white noise calculations of Refs. \cite{Tay}
and \cite{Bal}, cf. Chap.11, formula  (11.25). In particular, the
so-called running diffusion coefficient arises via straightforward
integration of the function $R_{11}(\tau )= <u_1(t)u_1(t_0)>$
where $\tau = t-t_0 >0$:

\begin{equation}
D_1(t) = \int_0^t <u_1(0)u_1(\tau)> d\tau  = %
{{k_BT}\over m} {{\beta + [\omega _csin(\omega _ct) - \beta
cos(\omega _ct)]exp(-\beta t)}\over {\beta ^2 +  {\omega _c}^2}}
\end{equation}
with an obvious asymptotics (the same for $D_2(t)$):
$D_B=lim_{t\rightarrow \infty } D_1(t)= {{k_BT}\over m} {\beta
\over {\beta ^2 + {\omega _c}^2}}$ and the large friction ($\omega
_c$ fixed and bounded) version $D= {{k_BT}\over {m\beta }}$.

\section{Spatial process}

The cylindrical symmetry of the problem allows us to consider
separately processes running on  the $XY$  plane  and along the
$Z$-axis (where the free Brownian motion takes place). We shall
confine further attention to the two-dimensional $XY$-plane
problem.
 Henceforth,  each vector will carry two components which
 correspond to the $x$ and $y$ coordinates respectively.
 We will directly refer to the
vector and matrix quantities introduced in the previous section,
but while keeping the same notation, we shall  simply disregard
their $z$-coordinate contributions.

We  define the spatial  displacement $\overrightarrow{r}$ of the
Brownian particle as folows

\begin{equation}
\overrightarrow{r}-\overrightarrow{r}_{0}=\int_{0}^{t}
\overrightarrow{u}%
\left( \eta \right) dn
\end{equation}

where $\overrightarrow{u}\left( t\right) $ is given by Eq.
(\ref{LanII}) (except for disregarding the third coordinate).

Our aim is to derive the probability distribution of
$\overrightarrow{r}$\ at time $t$ provided  that the particle
position and velocity were equal
  $\overrightarrow{r}_{0}$\ and
$\overrightarrow{u}_{0}$\ respectively, at time $t_{0}=0$.

To that end we shall mimic  procedures of the previous section. In
view of:

\begin{equation}
\overrightarrow{r}-\overrightarrow{r}_{0}-
\int_{0}^{t}e^{-\Lambda \eta }%
\overrightarrow{u}_{0}=\int_{0}^{t}d\eta
\int_{0}^{\eta }dse^{-\Lambda
\left( \eta -s\right) }\overrightarrow{A}\left( s\right)
\end{equation}

we have

\begin{equation}
\overrightarrow{r}-\overrightarrow{r}_{0}-
\Lambda ^{-1}\left( 1-e^{-\Lambda
t}\right) \overrightarrow{u}_{0}=\int_{0}^{t}\Lambda ^{-1}\left(
1-e^{\Lambda \left( s-t\right) }\right)
\overrightarrow{A}\left( s\right) ds
\end{equation}

where
\begin{equation}
\Lambda ^{-1}=\frac{1}{\beta ^{2}+\omega _{c}^{2}}\left(
\begin{array}{cc}
\beta & \omega _{c} \\
-\omega _{c} & \beta
\end{array}
\right)
\end{equation}

is the inverse  of  the matrix $\Lambda $ (regarded as a
rank two  sub-matrix   of that originally introduced in
 Eq. (3)).
Let us define  two auxiliary matrices

\begin{eqnarray}
\Omega &\equiv &\Lambda ^{-1}\left( 1-e^{-\Lambda t}\right)
 \label{omega} \\
\phi \left( s\right) &\equiv &\Lambda ^{-1}\left( 1-e^{\Lambda
\left( s-t\right) }\right)  \notag \enspace .
\end{eqnarray}

Because of:

\begin{equation}
e^{-\Lambda t}=\exp \left\{ - t \left(
\begin{array}{cc}
\beta & -\omega _{c} \\
\omega _{c} & \beta
\end{array}
\right) \right\} =e^{-\beta t}\left(
\begin{array}{cc}
\cos \omega _{c}t & \sin \omega _{c}t \\
-\sin \omega _{c}t & \cos \omega _{c}t
\end{array}
\right) =e^{-\beta t}U\left( t\right)
\end{equation}

we can represent  matrices  $\Omega $,  $\phi \left( s\right) $ in
more detailed   form. We have:

\begin{equation}
\Omega =\frac{1}{\beta ^{2}+\omega
_{c}^{2}}\left\{ \left(
\begin{array}{cc}
\beta & \omega _{c} \\
-\omega _{c} & \beta
\end{array}
\right) -e^{-\beta t}\left(
\begin{array}{cc}
\beta & \omega _{c} \\
-\omega _{c} & \beta
\end{array}
\right) \left(
\begin{array}{cc}
\cos \omega _{c}t & \sin \omega _{c}t \\
-\sin \omega _{c}t & \cos \omega _{c}t
\end{array}
\right) \right\}
\end{equation}

and

\begin{equation}
\phi \left( s\right) =\Lambda ^{-1}\left( 1-e^{-\beta \left(
t-s\right) }U\left( t-s\right) \right) =
\end{equation}
\begin{equation*}
\frac{1}{\beta ^{2}+\omega _{c}^{2}}\left(
\begin{array}{cc}
\beta & \omega _{c} \\
-\omega _{c} & \beta
\end{array}
\right) \left(
\begin{array}{cc}
1-e^{\beta \left( s-t\right) }\cos \omega _{c}\left( s-t\right) &
-e^{\beta
\left( s-t\right) }\sin \omega _{c}\left( s-t\right) \\
e^{\beta \left( s-t\right) }\sin \omega _{c}\left( s-t\right) &
1-e^{\beta
\left( s-t\right) }\cos \omega _{c}\left( s-t\right)
\end{array}
\right) \enspace .
\end{equation*}

Next steps imitate procedures of the previous section. Thus, we
seek for the probability distribution of the random (planar)
vector $ \overrightarrow{R}=\int_{0}^{t}\phi \left( s\right)
\overrightarrow{A}\left( s\right) ds  \label{Rdef}$ where
$\overrightarrow{R}=\overrightarrow{r}-\overrightarrow{r}_{0}-\Omega
\overrightarrow{u}_{0}$.

Dividing the time interval $\left( 0,t\right) $\ into  small
subintervals  to  assure that $\phi \left( s\right) $ can be
regarded  constant over the time span  $\left( j\Delta t,\left(
j+1\right) \Delta t\right) $\ and equal  $\phi \left( j\Delta
t\right) $,  we obtain

\begin{equation}
\overrightarrow{R}=\sum_{j=0}^{N-1}\phi \left( j\Delta t\right)
\int_{j\Delta t}^{\left( j+1\right) \Delta
t}\overrightarrow{A}\left(
s\right) ds=\sum_{j=0}^{N-1}\phi \left( j\Delta t\right)
\overrightarrow{B}%
\left( \Delta t\right) =\sum_{j=0}^{N-1}\overrightarrow{r}_{j}
\end{equation}

where  $\overrightarrow{r}_{j}=\phi \left( j\Delta t\right)
\overrightarrow{B}\left( \Delta t\right) =\phi
_{j}\overrightarrow{B}\left( \Delta t\right) $.

By invoking the probability distribution (10) we  perform an
appropriate change of variables: $\overrightarrow{r%
}_{j}=\phi _{j}\overrightarrow{B}\left( \Delta t\right) $ to yield
a probability  distribution of  $\overrightarrow{r}_{j}$

\begin{equation}
\widetilde{w}\left[ \overrightarrow{r}_{j}\right] =
\det \left[ \phi _{j}^{-1}%
\right] w\left[ \phi _{j}^{-1}\overrightarrow{r}_{j}\right]
=\frac{1}{\det \phi _{j}}w\left[ \phi
_{j}^{-1}\overrightarrow{r}_{j}\right] \enspace .
\end{equation}

Presently (not to be confused with previous steps (11)-(15)) we
have

\begin{equation}
\det \phi \left( s\right) =\frac{1}{\beta ^{2}+\omega
_{c}^{2}}\left( 1+e^{2\beta \left( s-t\right) }-2e^{\beta \left(
s-t\right) }\cos \omega _{c}\left( s-t\right) \right)
\end{equation}

and

\begin{equation}
\phi ^{-1}\left( s\right) =\frac{1}{1+e^{2\beta \left( s-t\right)
}-2e^{\beta \left( s-t\right) }\cos \omega _{c}\left( s-t\right)
}\left[ 1-e^{\beta \left( s-t\right) }U\left( -\left( s-t\right)
\right) \right] \Lambda  \enspace .
\end{equation}

So,  the inverse of the matrix  $\phi _{j}$ has the form:

\begin{equation}
\phi _{j}^{-1}=\frac{\widetilde{A}_{j}}{\gamma _{j}}
\end{equation}

where

\begin{equation}
\widetilde{A}_{j}=\left(
\begin{array}{cc}
1-e^{\beta \left( j\Delta t-t\right) }\cos \omega _{c}\left( j\Delta
t-t\right) & e^{\beta \left( j\Delta t-t\right) }\sin \omega _{c}\left(
j\Delta t-t\right) \\
-e^{\beta \left( j\Delta t-t\right) }\sin \omega _{c}\left( j\Delta
t-t\right) & 1-e^{\beta \left( j\Delta t-t\right) }\cos
\omega _{c}\left(
j\Delta t-t\right)
\end{array}
\right) \left(
\begin{array}{cc}
\beta & -\omega _{c} \\
\omega _{c} & \beta
\end{array}
\right)
\end{equation}

and

\begin{equation}
\gamma _{j}=1+e^{2\beta \left( j\Delta t-t\right) }-2e^{\beta
\left( j\Delta t-t\right) }\cos \omega _{c}\left( j\Delta
t-t\right) \enspace .
\end{equation}

There holds:

\begin{equation}
\det \phi _{j}^{-1}=\left( \det \phi _{j}\right) ^{-1}=\left(
\beta ^{2}+\omega _{c}^{2}\right) \frac{1}{\gamma _{j}}
\end{equation}

and as a consequence  the probability distribution of
$\overrightarrow{r}_{j}$ becomes

\begin{equation}
\widetilde{w}\left[ \overrightarrow{r}_{j}\right] =
\frac{1}{\frac{1}{\beta
^{2}+\omega _{c}^{2}}\gamma _{j}}\left( \frac{1}{4\pi q\Delta t}
\right) \exp
\left\{ \frac{\left| \widetilde{A}_{j}\left(
\begin{array}{c}
r_{j}^{x} \\
r_{j}^{y}
\end{array}
\right) \right| ^{2}}{\gamma _{j}^{2}4q\Delta t}\right\} \enspace
.
\end{equation}

In view of

\begin{equation}
\left| \widetilde{A}_{j}\left(
\begin{array}{c}
r_{j}^{x} \\
r_{j}^{x}
\end{array}
\right) \right| ^{2}=\left( \beta ^{2}+
\omega _{c}^{2}\right) \gamma _{j}%
\left[ \left( r_{j}^{x}\right) ^{2}+
\left( r_{j}^{y}\right) ^{2}\right]
\end{equation}

 that finally leads to

\begin{equation}
\widetilde{w}\left[ \overrightarrow{r}_{j}\right] = \left(
\frac{\beta ^2 + \omega _{c}^2}{4\pi q\Delta t \gamma _{j}}\right)
\exp \left\{ -\frac{(\beta ^2 + \omega _{c}^2)\,  \left|
\overrightarrow{r}_{j}\right| ^{2}} {4q\Delta t \gamma
_{j}}\right\} \enspace .
\end{equation}

Since this probability distribution is Gaussian with mean zero
and variance $%
\sigma _{j}^{2}=$ $2q\Delta t\frac{1}{\beta ^{2}+\omega _{c}^{2}}
\gamma _{j}$, the random vector$\ \overrightarrow{R}$ as a sum of
independent random variables $\overrightarrow{r}_{j}$ has the
distribution

\begin{equation}
w\left( \overrightarrow{R}\right) =\frac{1}{2\pi \sum_{j}
\sigma _{j}^{2}}%
\exp \left( -\frac{R_{x}^{2}+R_{y}^{2}}{2\sum_{j} \sigma
_{j}^{2}}\right) \enspace .
\end{equation}

\begin{equation}
\sigma ^{2}=\sum_{j}\sigma _{j}^{2}=2q\sum_{j}\Delta
t\frac{1}{\beta ^{2}+\omega _{c}^{2}}\gamma _{j} \enspace .
\end{equation}

In the limit of $\Delta t\rightarrow 0$ we arrive at the
integral

\begin{equation}
\sigma ^{2}=2q\frac{1}{\beta ^{2}+\omega _{c}^{2}}
\int_{0}^{t}\gamma \left(
s\right) ds
\end{equation}

with $ \int_{0}^{t}\gamma \left( s\right) ds=t+  \Theta $, where

\begin{equation}
\Theta = \Theta (t) = \frac{1}{2\beta }\left( 1-e^{-2\beta
t}\right) -2\frac{1}{\beta ^{2}+\omega _{c}^{2}}\left[ \beta
+\left( \omega _{c}\sin \omega _{c}t-\beta \cos \omega
_{c}t\right) e^{-\beta t}\right]      \enspace .
\end{equation}

That gives rise to  an ultimate form of the transition probability
density of the spatial  displacement process:

\begin{equation}
P\left( \overrightarrow{r},t|\overrightarrow{r}_{0},t_{0}=0,
\overrightarrow{u%
}_{0}\right) =\frac{1}{4\pi \frac{k_{B}T}{m}\frac{\beta }
{\beta ^{2}+\omega
_{c}^{2}}\left( t+\Theta \right) }\exp \left( -
\frac{\left| \overrightarrow{r%
}-\overrightarrow{r}_{0}-\Omega \overrightarrow{u}_{0}
\right| ^{2}}{4\frac{%
k_{B}T}{m}\frac{\beta }{\beta ^{2}+\omega _{c}^{2}}
\left( t+\Theta \right) }%
\right)
\end{equation}

with  $\Omega =\Omega (t)$  defined in Eqs. (\ref{omega}), (27).
Notice that an asymptotic diffusion coefficient
$D_B=D{\beta^2\over {\beta ^2 + \omega ^2_c}}$ of Section 3 (cf.
Eq. (20)) appears here as a spatial dispersion - attenuation
signature (when $\omega _c$ grows up at $\beta $ fixed).

The spatial displacement process governed by the above transition
probability density surely is \it not \rm Markovian. That can be
checked by inspection:  the Chapman-Kolmogorov identity is not
valid, like in  the standard free Brownian motion example where
the Ornstein-Uhlenbeck process induced (sole) spatial dynamics is
non-Markovian as well.

\section{Phase-space process}

\subsection{Free Brownian motion, Kramers equation and
local conservation laws}

We take advantage of  the cylindrical symmetry of our problem, and
consider separately the (free) Brownian  dynamics in the direction
parallel to the magnetic field vector, e.g. along the $Z$-axis.

That amounts to invoking a  familiar Ornstein-Uhlenbeck process
(in velocity/momentum) in its extended phase-space form. In the
absence of external forces, the kinetic (Kramers-Fokker-Planck
equation) reads:

\begin{equation}
 {\partial _t W + u\nabla _zW = \beta \nabla _u(Wu) + q
\triangle _uW} \end{equation}

where $q=D\beta ^2$.   Here  $\beta $ is the friction coefficient,
$D$ will be identified later with the spatial diffusion constant,
and (as before) we set $D=k_BT/m\beta $ in conformity with  the
Einstein fluctuation-dissipation identity.

The  joint probability distribution (in fact, density) $W(z,u,t)$
for a freely moving Brownian particle which at $t=0$ initiates its
motion at $x_0$ with an arbitrary inital velocity $u_0$ can  be
given in the form of the maximally symmetric displacement
probability law:

\begin{equation}
W(z,u,t)= W(R,S) = [4\pi ^2(FG-H^2)]^{-1/2} \cdot  exp\{ - {{GR^2
- HRS + FS^2}\over {2(FG - H^2)}}\}
\end{equation}

where $R=z-u_0(1-e^{-\beta t})\beta ^{-1}$, $S=u-u_0e^{-\beta t}$
while $ F = {D\over \beta }(2\beta t - 3 +4e^{-\beta t}-
e^{-2\beta t})$\, $G=D\beta (1-e^{-2\beta t})$ and
$H=D(1-e^{-\beta t})^2$.

For future reference, let us notice that  marginal  probablity
densities, in the Smoluchowski regime (take for granted  that time
scales $\beta ^{-1}$ and space scales $(D\beta ^{-1})^{1/2}$ are
irrelevant \cite{Cha}) display   familiar forms of the
Maxwell-Boltzmann probability density $w(u,t)=({m\over {2\pi
kT}})^{1/2}\, exp( - {{mu^2}\over {2k_BT}})$ and the diffusion
kernel $w(z,t) = (4\pi Dt)^{-1/2} exp( - {z^2\over {4Dt}})$
respectively.

A direct evaluation of the first and second local moment of the
phase-space probability density gives

\begin{equation}
 <u> = \int du\,
uW(z,u,t)= w(R) [(H/F)R + u_0e^{-\beta t}]
\end{equation}

 \begin{equation}
<u^2> = \int du \, u^2W(z,u,t) = ({{FG-H^2}\over F}+ {H^2\over
F^2}R^2) \cdot (2\pi F)^{-1/2} exp(- {R^2\over {2F}}) \enspace .
\end{equation}

Let us  notice that  after  passing to the diffusion
(Smoluchowski) regime, \cite{Gar1},
 one   readily recovers
the  local (configuration space conditioned) moment
 $<u>_z={1\over w}<u>$ to be in the form

\begin{equation}
 <u>_z = {z\over {2t}}= - D{{\nabla w(z,t)}\over {w(z,t)}}
\end{equation}

 while for  the second local moment $<u^2>_z= {1\over w}<u^2>$
 we would arrive at

\begin{equation}
 {<u^2>_z = (D\beta - D/2t) + <u>^2_z \enspace .}
 \end{equation}

By inspection one verifies that the transport (Kramers)
 equation  for $W(z,u,t)$ implies local conservation laws:

 \begin{equation}
  \partial _t w + \nabla (<u>_z w) = 0
  \end{equation}

 and

 \begin{equation}
  {\partial _t(<u>_z w) + \nabla _z(<u^2>_zw) = - \beta <u>_z w
 \enspace .}
 \end{equation}

 At this point  (we strictly follow the moment
equations strategy of the traditional kinetic theory of gases and
liquids, compare e.g. \cite{Lag}) let us introduce the notion of
the pressure function $P_{kin}$:

\begin{equation}
 P_{kin}(z,t) = (<u^2>_z - <u>^2_z) w(z,t)
 \end{equation}

in terms of which   we can analyze
the local momentum conservation law

\begin{equation}
 (\partial _t + <u>_z \nabla )<u>_z = - \beta <u>_z -
{{\nabla P_{kin}} \over w} \enspace .
\end{equation}

One should realize that in the Smoluchowski regime  the
friction term is cancelled away by
a counterterm  coming from ${1\over w} \nabla P_{kin}$  so that

\begin{equation}
 (\partial _t + <u>_z\nabla )<u>_z = {D\over {2t}}{{\nabla
w}\over w} = - {{\nabla P}\over w}
\end{equation}

where $P=D^2w\triangle ln\, w$, called osmotic pressure in Ref.
\cite{Nel}, is the net remnant of the kinetic pressure
contribution.

Further exploiting the kinetic lore, we can   tell  few words
about the \it  temperature of Brownian particles \rm as opposed to
the (equilibrium) temperature of the thermal bath. Namely, in view
of  (we refer  to  the Smoluchowski regime with $t\geq \beta ^{-1}$)
$P_{kin} \sim (D\beta
- {D\over {2t}})w$ where $D={{k_BT}\over {m\beta }}$, we can
formally  set:

 \begin{equation}
  { k_B T_{kin} = {P_{kin} \over w} \sim (k_BT - {D\over {2t}}) <
k_BT }\enspace .
\end{equation}

That quantifies the degree of thermal agitation (temperature) of
Brownian particles to be \it less \rm than the thermostat
temperature. Heat is continually pumped from the  thermostat to
the Brownian "gas", until asymptotically both temperatures
equalize. This  may  be called a "thermalization" of Brownian
particles. In the process of that "thermalization" the   Brownian
"gas" temperature monotonically  grows up until  the mean kinetic
energy of particles     and that of mean flows asymptotically
approach the familiar kinetic relationship: $ \int  {w\over
2}(<u^2>_z - <u>^2_z) dx = k_BT$,  cf. Refs. \cite{Gar},
\cite{Gar1} for more extended discussion of that medium
$\rightarrow $ particles heat transfer issue and its possible
relevance while associating habitual thermal equilibrium
conditions with essentially non-equilibrium phenomena.

{\bf Remark 1:} Once local conservation laws were introduced, it
seems instructive to comment on the essentially hydrodynamical
features (compressible fluid/gas case) of the problem.
Specifically,  the "pressure" term $\nabla Q$ is  quite annoying
from the traditional kinetic theory perspective. That is  apart
from the fact that our local conservation laws have a conspicuous
Euler form appropriate for the standard hydrodynamics of gases and
liquids. One should become alert that in the present (Brownian)
context they convey an entirely different message. For example, in
case of normal liquids the pressure is exerted upon any control
volume (droplet) by the surrounding fluid. We may interpret that
as a compression of a droplet. In case of Brownian motion, we deal
with a definite decompression: particles are driven away from
areas of higher concentration (probability of occurence). Hence,
typically the Brownian "pressure" is exerted by the droplet upon
its surrounding.

{\bf Remark 2:}
 The  derivation of a hierarchy of local conservation laws
(moment equations)  for the Kramers equation can be patterned
after the standard procedure for the Boltzmann equation. Those
laws do not form a closed system and additional specifications
(like the familiar thermodynamic equation of state) are needed to
that end. In case of the isothermal Brownian motion, when
considered in the large friction regime (e.g. Smoluchowski
diffusion approximation), it suffices to supplement  the
Fokker-Planck equation  by one more conservation law \it only \rm
to arrive at  a closed system, \cite{Gar} and compare with the
discussion of Ref. \cite{Lag}.

\subsection{Planar process}

Now we shall consider  Brownian dynamics in the direction
perpendicular to the magnetic  field $\overrightarrow{B}$, hence
(while in terms of configuration-space variables) we address an
issue of  the planar dynamics. We are  interested in the complete
phase-space process, hence we need  to specify the transition
probability density \ $P\left(
\overrightarrow{r},\overrightarrow{u},t|\overrightarrow{r}_{0},%
\overrightarrow{u}_{0},t_{0}=0\right) $ of the Markov
 process conditioned by the initial  data
 $\overrightarrow{u}=%
\overrightarrow{u}_{0}$ and $\overrightarrow{r}=
\overrightarrow{r}_{0}$ at time $%
t_{0}=0$. That  is equivalent to deducing the joint probability
distribution
$W\left( \overrightarrow{S,}%
\overrightarrow{R}\right) $\ of random vectors
$\overrightarrow{S}$ and $%
\overrightarrow{R}$, previously defined
to appear in the form $\overrightarrow{S}=\overrightarrow{u}
\left( t\right) -e^{-\Lambda t}%
\overrightarrow{u}_{0}$ and
$\overrightarrow{R}=\overrightarrow{r}-\overrightarrow{r}_{0}-
\Omega \overrightarrow{u}_{0}$   respectively, cf. Eqs. (15) and (44).

Let us stress that presently, all vectors are regarded as
two-dimensional versions (the third component being simply
disregarded) of the original random variables we have discussed so
far in Sections 2 and 3.

 Vectors  $\overrightarrow{S}$ and
$\overrightarrow{R}$ both share
 a Gaussian distribution with mean zero. Consequently,
  the joint distribution
$W\left( \overrightarrow{S,}\overrightarrow{R}\right) $\ is
determined by the matrix of variances and covariances: $C = \left(
c_{ij}\right) = \left(\left\langle x_{i}x_{j}\right\rangle
\right)$, where  we abbreviate four phase-space variables in a
single notion of  $x=\left( S_{1},S_{2},R_{1},R_{2}\right) $ and
label components of $x$ by $i,j=1,2,3,4$.  In terms of
$\overrightarrow{R}$ and $\overrightarrow{S}$ the covariance
matrix $C$ reads:

\begin{equation}
C=\left(
\begin{array}{cccc}
\left\langle S_{1}S_{1}\right\rangle &
\left\langle S_{1}S_{2}\right\rangle
& \left\langle S_{1}R_{1}\right\rangle &
\left\langle S_{1}R_{2}\right\rangle
\\
\left\langle S_{2}S_{1}\right\rangle &
\left\langle S_{2}S_{2}\right\rangle
& \left\langle S_{2}R_{1}\right\rangle &
\left\langle S_{2}R_{2}\right\rangle
\\
\left\langle R_{1}S_{1}\right\rangle &
\left\langle R_{1}S_{2}\right\rangle
& \left\langle R_{1}R_{1}\right\rangle &
\left\langle R_{1}R_{2}\right\rangle
\\
\left\langle R_{2}S_{1}\right\rangle &
\left\langle R_{2}S_{2}\right\rangle
& \left\langle R_{2}R_{1}\right\rangle &
\left\langle R_{2}R_{2}\right\rangle
\end{array}
\right) \enspace .
\end{equation}

The joint probability distribution of $\overrightarrow{S}$ and $%
\overrightarrow{R}$ is  given by

\begin{equation}
W\left( \overrightarrow{S,}\overrightarrow{R}\right) =W\left(
\overrightarrow{x}\right) =\frac{1}{4\pi ^{2}} \left(
\frac{1}{\det C}\right)^{\frac{1}{2}}\exp
\left(
-\frac{1}{2}\sum_{i,j}c_{ij}^{-1}x_{i}x_{j}\right)
\end{equation}

where $c_{ij}^{-1}$denotes the component of the inverse matrix
$C^{-1}$.

The probability distributions of $\overrightarrow{S}$ and
$\overrightarrow{R} $, which were established in the previous
sections, determine a number of expectation values:
\begin{equation}
g\equiv \left\langle S_{1}S_{1}\right\rangle =\left\langle
S_{2}S_{2}\right\rangle =\frac{k_{B}T}{m}\left( 1-e^{-2\beta t}\right)
\end{equation}

cf. Eq. (14), while $ \left\langle S_{1}S_{2}\right\rangle =
\left\langle
S_{2}S_{1}\right\rangle =0$. Furthermore:

\begin{equation}
f\equiv \left\langle R_{1}R_{1}\right\rangle =\left\langle
R_{2}R_{2}\right\rangle =2\frac{k_{B}T}{m}\frac{\beta } {\beta
^{2}+\omega _{c}^{2}}\left( t+\Theta \right)= 2D_B (t+\Theta )
\end{equation}

cf. Eqs. (20, (42), (43). In addition we have
 $ \left\langle R_{1}R_{2}\right\rangle
=\left\langle R_{2}R_{1}\right\rangle =0$.

As a  consequence, we are left with only   four non-vanishing
components of the covariance matrix $C$: $
c_{13}=c_{31}=\left\langle S_{1}R_{1}\right\rangle $, $
c_{14}=c_{41}=\left\langle S_{1}R_{2}\right\rangle $, $
c_{23}=c_{32}=\left\langle S_{2}R_{1}\right\rangle $,  $
c_{24}=c_{42}=\left\langle S_{2}R_{2}\right\rangle $ which need a
closer examination.

We can obtain those covariances by exploiting
a dependence of  the random quantities $%
\overrightarrow{S}$ and $\overrightarrow{R}$  on the white-noise
term $%
\overrightarrow{A}\left( s\right) $ whose statistical properties
are known. There follows:

\begin{equation}
S_{1}=\int_{0}^{t}dse^{-\beta \left( t-s\right) }\left[ \cos \omega
_{c}\left( t-s\right) A_{1}\left( s\right) +\sin \omega _{c}\left(
t-s\right) A_{2}\left( s\right) \right]
\end{equation}

\begin{equation*}
S_{2}=\int_{0}^{t}dse^{-\beta \left( t-s\right) }\left[ -\sin \omega
_{c}\left( t-s\right) A_{1}\left( s\right) +\cos \omega _{c}\left(
t-s\right) A_{2}\left( s\right) \right]
\end{equation*}

\begin{eqnarray*}
R_{1} &=&\int_{0}^{t}ds\frac{1}{\beta ^{2}+\omega _{c}^{2}}\left[
\beta \left( 1-e^{-\beta \left( t-s\right) }\cos \omega _{c}\left(
t-s\right) \right) +\omega _{c}e^{-\beta \left( t-s\right) }\sin
\omega _{c}\left( t-s\right) \right] A_{1}\left( s\right) + \\ &
&\int_{0}^{t}ds\frac{1}{\beta ^{2}+\omega _{c}^{2}}\left[ -\beta
e^{-\beta \left( t-s\right) }\sin \omega _{c}\left( t-s\right)
+\omega _{c}\left( 1-e^{-\beta \left( t-s\right) }\cos \omega _{c}
\left( t-s\right) \right) %
\right] A_{2}\left( s\right)
\end{eqnarray*}

\begin{eqnarray*}
R_{2} &=&\int_{0}^{t}ds\frac{1}{\beta ^{2}+\omega _{c}^{2}} \left[
-\omega _{c}\left( 1-e^{-\beta \left( t-s\right) } \cos \omega
_{c}\left( t-s\right) \right) +\beta e^{-\beta \left( t-s\right)
}\sin \omega _{c}\left( t-s\right) \right] A_{1}\left( s\right) +
\\ &&\int_{0}^{t}ds\frac{1}{\beta ^{2}+\omega _{c}^{2}}\left[
\omega _{c}e^{-\beta \left( t-s\right) }\sin \omega _{c} \left(
t-s\right) +\beta \left( 1-e^{-\beta \left( t-s\right) }\cos
\omega _{c} \left( t-s\right) \right) \right] A_{2}\left( s\right)
\enspace .
\end{eqnarray*}

Multiplying together suitable  components of  vectors
$\overrightarrow{S}$ and $\overrightarrow{R}$  and taking averages
of those products in conformity with
the rules $\left\langle A_{i}\left( s\right) \right\rangle =0$ and $%
\left\langle A_{i}\left( s\right) A_{j}\left( s^{\shortmid }\right)
\right\rangle =2q\delta _{ij}\delta \left( s-s^{\shortmid }\right) $,
where $%
q=\frac{k_{B}T}{m}\beta $, $i,j=1,2,3$, we arrive at:

\begin{equation}
h\equiv \left\langle R_{1}S_{1}\right\rangle = \left\langle
R_{2}S_{2}\right\rangle =2q\frac{1}{\beta ^{2}+ \omega _{c}^{2}}
\int_{0}^{t}ds e^{-\beta \left( t-s\right) } [ \beta \cos \omega
_{c}\left( t-s\right) +
\end{equation}
\begin{equation*}
\omega _{c}\sin \omega _{c} \left( t-s\right) -\beta e^{-\beta
\left( t-s\right) }]  =q\frac{1}{\beta ^{2}+\omega _{c}^{2}}
\left( 1-2e^{-\beta t}\cos \omega _{c}t+e^{-2\beta t}\right)
\end{equation*}

and
\begin{equation}
k\equiv \left\langle R_{1}S_{2}\right\rangle =-\left\langle
R_{2}S_{1}\right\rangle =2q\frac{1}{\beta ^{2}+\omega _{c}^{2}}%
\int_{0}^{t}dse^{-\beta \left( t-s\right) }[ -\beta \sin \omega
_{c}\left( t-s\right) +
\end{equation}
\begin{equation*}
\omega _{c}\cos \omega _{c} \left( t-s\right) -\omega
_{c}e^{-\beta \left( t-s\right) }] = q\frac{1}{\beta ^{2}+\omega
_{c}^{2}}\left[ 2e^{-\beta t}\sin \omega _{c}t-\frac{\omega
_{c}}{\beta }\left( 1-e^{-2\beta t}\right) \right] \enspace .
\end{equation*}

The covariance matrix $C=\left( c_{ij}\right) $ has thus the form

\begin{equation}
C=\left(
\begin{array}{cccc}
g & 0 & h & -k \\
0 & g & k & h \\
h & k & f & 0 \\
-k & h & 0 & f
\end{array}
\right)
\end{equation}

while its  inverse  $C^{-1}$ reads as follows:

\begin{equation}
C^{-1}=\frac{1}{\det C}\left( fg-h^{2}-k^{2}\right) \left(
\begin{array}{cccc}
f & 0 & -h & k \\
0 & f & -k & -h \\
-h & -k & g & 0 \\
k & -h & 0 & g
\end{array}
\right)
\end{equation}

where $ \det C=\left( fg-h^{2}-k^{2}\right) ^{2}$.

The joint probability distribution of $\overrightarrow{S}$ and $%
\overrightarrow{R}$\ can be ultimately written in the  form:

\begin{equation}
W\left( \overrightarrow{S},\overrightarrow{R}\right) =
\end{equation}
\begin{equation}
\frac{1}{4\pi ^{2}\left( fg-h^{2}-k^{2}\right) }\exp \left( -
\frac{f\left| \overrightarrow{%
S}\right| ^{2}+g\left| \overrightarrow{R}\right| ^{2}-
2h\overrightarrow{S}%
\cdot \overrightarrow{R}+2k\left( \overrightarrow{S}
\times \overrightarrow{R}%
\right) _{i=3}}{2\left( fg-h^{2}-k^{2}\right) }\right) \enspace .
\end{equation}

In the above, all vector entries are two-dimensional. The specific
 $i=3$ vector product coordinate in the exponent is simply an
abbreviation for the (ordinary $R^3$-vector product) procedure
that involves merely  first two components of three-dimensional
vectors (the third is then arbitrary and irrelevant), hence
effectively involves  our two-dimensional $\overrightarrow{R}$ and
$\overrightarrow{S}$.

\subsection{Kramers equation and local conservation laws for
the planar motion}

For the purpose of evaluating local velocity averages of the
Kramers equation, we need to extract the marginal configuration
space distribution.  Let us notice that

\begin{equation}
\int W(\overrightarrow{S},\overrightarrow{R})d\overrightarrow{S} =
w(\overrightarrow{R})
= P(\overrightarrow{r},t|\overrightarrow{r}_0,t_0=0,
\overrightarrow{u}_0)
\end{equation}

where the last transition probability  density entry coincides
with that of Eq. (44).

Let us introduce an auxiliary (weighted) distribution:

\begin{equation}
\widetilde{W}\left( \overrightarrow{S}|\overrightarrow{R}\right) =
\frac{%
W\left( \overrightarrow{S},\overrightarrow{R}\right) }{\int W\left(
\overrightarrow{S},\overrightarrow{R}\right) d\overrightarrow{S}} =
\end{equation}
\begin{equation*}
\frac{1}{%
2\pi \frac{1}{f}\left( fg-h^{2}-k^{2}\right) }\exp
\left( -\frac{\left|
\overrightarrow{S}-\overrightarrow{m}
\right| ^{2}}{2\frac{1}{f}\left(
fg-h^{2}-k^{2}\right) }\right)
\end{equation*}

where
\begin{equation}
\overrightarrow{m}=\frac{1}{f}\left( hR_{1}-kR_{2},hR_{2}+
kR_{1}\right)
\end{equation}

and we recall that $
\overrightarrow{S} = \overrightarrow{u}\left( t\right) -
e^{-\beta t}U\left(
t\right) \overrightarrow{u}_{0}$ and
 $\overrightarrow{R} = \overrightarrow{r}-
 \overrightarrow{r}_{0}-\Omega
\overrightarrow{u}_{0}$.

The local expectation values (compare e.g. calculations of the
previous subsection) read:
$\left\langle u_{i}\right\rangle _{\overrightarrow{R}} = \int u_{i}%
\widetilde{W}d\overrightarrow{u} $ and
$ \left\langle u_{i}^{2}\right\rangle _{\overrightarrow{R}} =
\int u_{i}^{2}%
\widetilde{W}d\overrightarrow{u}$ where $i=1,2$.

By evaluating those averages we get:

\begin{equation}
\left\langle \overrightarrow{u}\right\rangle _{\overrightarrow{R}}
= \left( \left\langle u_{1}\right\rangle
_{\overrightarrow{R}},\left\langle u_{2}\right\rangle
_{\overrightarrow{R}}\right)
 = e^{-\beta t}U\left( t\right) \overrightarrow{u}_{0} +
\overrightarrow{m}
\end{equation}

and

\begin{equation}
\left\langle u_{1}^{2}\right\rangle _{\overrightarrow{R}} -
\left\langle
u_{1}\right\rangle _{\overrightarrow{R}}^{2} =
\left\langle u_{2}^{2}\right\rangle _{\overrightarrow{R}} -
\left\langle
u_{2}\right\rangle _{\overrightarrow{R}}^{2}  =
\frac{1}{f}\left(fg-h^{2}-k^{2}\right)
\end{equation}

The Fokker-Planck-Kramers  equation, appropriate for the planar
dynamics in its  phase space version, reads

\begin{equation}
\frac{\partial W}{\partial t}+\overrightarrow{u}
\nabla _{\overrightarrow{r}%
}W=\beta \nabla _{\overrightarrow{u}}
\left( W\overrightarrow{u}\right)
-\omega _{c}\left[ \nabla _{\overrightarrow{u}}\times
W\overrightarrow{u}%
\right] _{i=3}+q\nabla _{\overrightarrow{u}}^{2}W
\end{equation}

where the troublesome again  (in the planar case all vectors are
two-dimensional)  vector product third component stands for
\begin{equation}
\left[ \nabla _{\overrightarrow{u}}\times
W\overrightarrow{u}\right] _{i=3}=%
\frac{\partial }{\partial u_{1}}\left( Wu_{2}\right) -
\frac{\partial }{%
\partial u_{2}}\left( Wu_{1}\right) \enspace .
\end{equation}

First two moment equations are easily derivable. Namely, the
continuity (0-th moment) and the momentum conservation (first
moment) equations come out in the form

\begin{equation}
\partial _{t}w+\overrightarrow{\nabla } \cdot \left[ \left\langle
\overrightarrow{u}\right\rangle _{%
\overrightarrow{R}}w\right] =0
\end{equation}

and

\begin{equation}
\partial _{t}\left[ \left\langle u_{1}
\right\rangle _{\overrightarrow{R}}w%
\right] +\frac{\partial }{\partial r_{1}}\left[ \left\langle
u_{1}^{2}\right\rangle _{\overrightarrow{R}}w\right] +
\frac{\partial }{%
\partial r_{2}}\left[ \left\langle u_{1}
\right\rangle _{\overrightarrow{R}%
}\left\langle u_{2}\right\rangle _{
\overrightarrow{R}}w\right] =-\beta
\left\langle u_{1}\right\rangle _{\overrightarrow{R}}w+\omega
_{c}\left\langle u_{2}\right\rangle _{\overrightarrow{R}}w
\end{equation}
\begin{equation*}
\partial _{t}\left[ \left\langle u_{2}
\right\rangle _{\overrightarrow{R}}w%
\right] +\frac{\partial }{\partial r_{2}}\left[ \left\langle
u_{2}^{2}\right\rangle _{\overrightarrow{R}}w\right] +
\frac{\partial }{%
\partial r_{1}}\left[ \left\langle u_{1}
\right\rangle _{\overrightarrow{R}%
}\left\langle u_{2}\right\rangle _{\overrightarrow{R}}w\right]
 =-\beta
\left\langle u_{2}\right\rangle _{\overrightarrow{R}}w-\omega
_{c}\left\langle u_{1}\right\rangle _{\overrightarrow{R}}w
\enspace .
\end{equation*}

That implies

\begin{equation}
\left[ \partial _{t}+\left\langle u_{1}\right\rangle _{
\overrightarrow{R}}%
\frac{\partial }{\partial r_{1}}+\left\langle u_{2}\right\rangle _{%
\overrightarrow{R}}\frac{\partial }{\partial r_{2}}\right]
\left\langle
u_{1}\right\rangle _{\overrightarrow{R}}=-\beta \left\langle
u_{1}\right\rangle _{\overrightarrow{R}}+\omega _{c}\left\langle
u_{2}\right\rangle _{\overrightarrow{R}}-\frac{1}{w}\frac{\partial }{%
\partial r_{1}}\left[ \left( \left\langle u_{1}^{2}\right\rangle _{%
\overrightarrow{R}}-\left\langle u_{1}
\right\rangle _{\overrightarrow{R}%
}^{2}\right) w\right]
\end{equation}
\begin{equation*}
\left[ \partial _{t}+\left\langle u_{1}
\right\rangle _{\overrightarrow{R}}%
\frac{\partial }{\partial r_{1}}+\left\langle u_{2}\right\rangle _{%
\overrightarrow{R}}\frac{\partial }{\partial r_{2}}
\right] \left\langle
u_{2}\right\rangle _{\overrightarrow{R}}=-\beta \left\langle
u_{2}\right\rangle _{\overrightarrow{R}}-\omega _{c}\left\langle
u_{1}\right\rangle _{\overrightarrow{R}}-\frac{1}{w}\frac{\partial }{%
\partial r_{2}}\left[ \left( \left\langle u_{2}^{2}\right\rangle _{%
\overrightarrow{R}}-\left\langle u_{2}
\right\rangle _{\overrightarrow{R}%
}^{2}\right) w\right]
\end{equation*}

which finally sums up to a local momentum conservation law (here,
the standard $R^3$  vector product on the right-hand-side
contributes its first and second components only)

\begin{equation}
\left[ \partial _{t}+\left\langle \overrightarrow{u}
\right\rangle _{%
\overrightarrow{R}}\overrightarrow{\nabla }\right]
\left\langle \overrightarrow{u}%
\right\rangle _{\overrightarrow{R}} =
-\Lambda \left\langle \overrightarrow{%
u}\right\rangle _{\overrightarrow{R}}-\frac{1}{w}
\overrightarrow{\nabla }\cdot \overleftrightarrow{P}_{kin}= -\beta
\left\langle \overrightarrow{u}
\right\rangle _{\overrightarrow{R}}+%
\frac{q_{e}}{mc}\left\langle \overrightarrow{u}
\right\rangle _{%
\overrightarrow{R}} \times
\overrightarrow{B}-\frac{1}{w}\overrightarrow{\nabla }\cdot
\overleftrightarrow{P}_{kin}
\end{equation}

where $\overleftrightarrow{P}_{kin}$ has tensor  components $%
P^{kin}_{ij}$, and $\overrightarrow{\nabla }\cdot
\overleftrightarrow{P}_{kin}$ stands for  a vector whose  $i$-th
component is equal
 $\sum_{j}\frac{\partial P^{kin}_{ij}}{\partial r_{j}}$  and
$i,j=1,2 $.  Here, obviously $P^{kin}_{ij}=\left\langle \left(
u_{i}-\left\langle u_{i}
\right\rangle _{%
\overrightarrow{R}}\right) \left( u_{j}-
\left\langle u_{j}\right\rangle _{%
\overrightarrow{R}}\right) \right\rangle _{\overrightarrow{R}}w $
and only diagonal entries  do not vanish. Clearly $ P^{kin}_{ii}
=\left( \left\langle u_{i}^{2}
\right\rangle _{\overrightarrow{R}%
}-\left\langle u_{i}
\right\rangle _{\overrightarrow{R}}^{2}\right) w $.

Because of

\begin{equation}
\sigma ^{2}=\left\langle u_{1}^{2}\right\rangle _{\overrightarrow{R}%
}-\left\langle u_{1}
\right\rangle _{\overrightarrow{R}}^{2}=\left\langle
u_{2}^{2}\right\rangle _{\overrightarrow{R}}-
\left\langle u_{2}\right\rangle
_{\overrightarrow{R}}^{2}=
\frac{1}{f}\left( fg-h^{2}-k^{2}\right) =g-\frac{%
h^{2}+k^{2}}{f}
\end{equation}

we can introduce again  $\sigma ^2= {P_{kin}\over w}= k_BT_{kin}$
and pass to an asymptotic regime $t>>t_{c}=\frac{1}{\beta }$.
Then,  we obtain Eq. (56) to be valid in the present case as well
thus  quantifying    an overall (magnetic field independent)
heating process involved .

In that asymptotic  regime we have $ \sigma ^{2}=D\beta
-\frac{D}{2t}$ and  by employing an asymptotic form  of
$w(\overrightarrow{R})$, Eq. (44) we recover:

\begin{equation}
\overrightarrow{\nabla }\cdot \overleftrightarrow{P}_{kin}=
\left( D\beta -\frac{D}{2t}\right) \frac{%
\overrightarrow{\nabla }w}{w}
\end{equation}

together with

\begin{equation}
\Lambda \left\langle
\overrightarrow{u}\right\rangle _{\overrightarrow{R}%
}=-D\beta \frac{\overrightarrow{\nabla }w}{w}   \enspace .
\end{equation}

 So,  asymptotically ($t>> \beta ^{-1}$)  the momentum
 conservation law
takes the form (to be compared with considerations of section
$4.1$)

\begin{equation}
\left[ \partial _{t}+\left\langle \overrightarrow{u}\right\rangle _{%
\overrightarrow{R}}\overrightarrow{\nabla }\right]
\left\langle \overrightarrow{u}%
\right\rangle
_{\overrightarrow{R}}=\frac{D}{2t}\frac{\overrightarrow{\nabla }
w}{w}
\end{equation}

However an asymptotic regime does not yet imply that the
right-hand-side of Eq. (83) represents an acceptable  "osmotic
pressure" gradient contribution. We additionally need a large
friction regime to deal with a consistent picture of a Markov
diffusion process in the Smoluchowski form. Indeed, to reproduce a
universal (Ref. \cite{Gar}) pressure-type functional dependence on
 $P$  we must employ  a suitable form of the diffusion
coefficient. The usage of $D$ alone to define an "osmotic
pressure" implies an apparent  failure. On the other hand, the
usage of $D_B = D \frac{\beta ^2}{\beta ^2 + {\omega ^2}_c}$ as
suggested by Eq. (44) leads to:

\begin{equation}
P=D_{B}^{2}\Delta \ln w \Rightarrow -\frac{\overrightarrow{\nabla
}P}{w} =\frac{D_{B}}{2t}\frac{\overrightarrow{\nabla }w}{w}
\enspace .
\end{equation}

Then however the momentum conservation law displays a
supplementary  scaling of the osmotic pressure contribution, which
may trivialize (die out) only for large values of $\beta $ (an
ultimate Smoluchowski regime). Namely, we have:

\begin{equation}
\left[ \partial _{t}+\left\langle \overrightarrow{u}
\right\rangle _{%
\overrightarrow{R}}\overrightarrow{\nabla }\right]
\left\langle \overrightarrow{u}%
\right\rangle _{\overrightarrow{R}}=-\left( \frac{\beta
^{2}}{\beta ^{2}+\omega _{c}^{2}}\right)
^{-1}\frac{\overrightarrow{\nabla }P}{w} \enspace .
\end{equation}

Such process is yet non-Markovian and its approximation by the
Smoluchowski process becomes  reliable when $\beta $ is large
while $\omega _c$ is kept moderately small.

Basically, the large friction regime cancels all rotational
features (arising due to the Lorentz force) on very short time
scales. If we are satisfied with the non-Markovian regime of
moderate friction but arbitrarily varying $\omega _c$ (i.e. $B$)
then, in conformity with Eq. (78),  mean flows would  display
signatures of rotation  that is bound to die out asymptotically.
This effect can be analyzed in $R^3$ by observing that a
three-dimensional extension of the vector $\overrightarrow{m}$ of
Eqs. (70), (71) asymptotically reads $\overrightarrow{m} =
\frac{1}{2t\beta }\Lambda
(\overrightarrow{r}-\overrightarrow{r}_0)$ where $\Lambda $ comes
from Eq. (3) and $(\overrightarrow{r}-\overrightarrow{r}_0) \in
R^3$. In view of that, we have (cf. Eq. (71)) $curl \left\langle
\overrightarrow{u}\right\rangle _{\overrightarrow{R}}\sim curl
\overrightarrow{m} \sim (0,0,-\frac{\omega _c}{t\beta })$ and the
circulation asymptotically vanishes. The effect can be slowed down
by a suitable adjustment of  $\omega _c$ against $\beta $.

{\bf Acknowledgement:} One of the authors (P. G.) receives
financial support from the KBN research grant No. 2 PO3B 086 16.

\end{document}